\documentclass[%
 aip,
 amsmath,amssymb,
reprint, onecolumn 
]{revtex4-1}

\usepackage{booktabs}
\usepackage{multirow}
\usepackage{threeparttable}
\usepackage{graphicx}
\usepackage{dcolumn}
\usepackage{bm}
\usepackage[utf8]{inputenc}
\usepackage[T1]{fontenc}
\usepackage{mathptmx}
\usepackage{etoolbox}
\usepackage{color}
\usepackage[margin=1.8cm]{geometry}
\usepackage[normalem]{ulem}
\usepackage{url}

\begin{document}

\title{Minimizing propagated density errors of atomic core-electron for simultaneously accurate bandgaps and lattice constants in closed-shell Copper semiconductors}

\author{Kuiyu Ye}
\affiliation {Key Lab of advanced optoelectronic quantum architecture and measurement (MOE), and School of Interdisciplinary Science, Beijing Institute of Technology, Beijing 100081, China}
\author{Haitao Liu}
\affiliation {Institute of Applied Physics and Computational Mathematics, Beijing 100088, China}
\affiliation {National Key Laboratory of Computational Physics, Beijing 100088, China}
\author{Yuanchang Li}
\email{yuancli@bit.edu.cn}
\affiliation {Key Lab of advanced optoelectronic quantum architecture and measurement (MOE), and School of Interdisciplinary Science, Beijing Institute of Technology, Beijing 100081, China}
\author{Shengbai Zhang}
\affiliation {Department of Physics, Applied Physics, and Astronomy, Rensselaer Polytechnic Institute, Troy, New York 12180, USA}

\date{\today}

\begin{abstract}
Density functional theory struggles to accurately determine electron density of atoms, whose error is inevitably encoded into the pseudopotential and propagated into solid-state calculations. However, little is known about how this affects accuracy nor how to remedy it. In this work, through a systematic study of the effect of Cu atomic density on bandgap and lattice constants of over 50 Cu-containing simple closed-shell semiconductors, we find that core-electron density can drastically affect nuclear attraction to valence electrons and subsequent charge distribution and energy position of Cu 3$d$ electrons. The error can be eliminated at its source by employing modified Hartree-Fock pseudopotentials for Cu core while retaining (semi-)local functionals for valence electrons. This real-space partitioning approach leads to simultaneous high-accuracy in bandgap and lattice constants across the entire material class.

\end{abstract}

\maketitle

\section{INTRODUCTION}
Density functional theory (DFT) stands as the workhorse of modern electronic structure calculations \cite{jones_density_2015,huang_central_2023}. With never-ending upgrades in hardware resources and innovation in theoretical methods, DFT has evolved from pursuing on-the-appearance correct results to increasingly emphasizing correct results with correct reasons \cite{hammes-schiffer_conundrum_2017}. Hohenberg-Kohn theorem laid the foundation for DFT, by providing rigorous mathematical proof that electron density distribution uniquely determines all properties of a system's ground state, and this density can be obtained through the minimization of a hypothetical exact energy functional \cite{hohenberg_inhomogeneous_1964,kohn_self-consistent_1965}. Despite the tremendous success of various approximations to the exact functional, whose academic influence ranks among the top across all sciences \cite{van_noorden_top_2014}, they all suffer from unavoidable systematic errors due to violations of certain mathematical constraints inherent to the exact functional.

Work on atomic species \cite{medvedev_density_2017} reveals that even functionals that accurately predict total energy can yield incorrect electron density, revealing that DFT can produce correct results for the wrong reason \cite{graziano_quantum_2017}. However, little research exists on how such atomic density errors affect the accuracy of solid-state DFT. Although solids are composed of atoms, their physical properties are primarily determined by the collective behavior of outer-shell electrons. As such, understanding solids through a homogeneous electron gas model is often more effective than an atomic picture \cite{loos_uniform_2016}. This distinction between solids and atoms is strikingly illustrated by the divergent performance of the same functional: one that excels in atoms fails in solids \cite{perdew_perdew_1998}, while one that excels in solids fails in atoms \cite{zhang_comment_1998}. By first principles, experimentally measurable properties are manifestations of microscopic electronic structure, not the other way around. In other words, errors in the atomic density will inevitably affect the accuracy of solid-state DFT, one way or the other.

Solid-state DFT widely employs pseudopotentials \cite{hellmann_new_1935,cohen_fifty_1983,ye_pseudopotentials_2025}, which distinguish core-electrons from valence-electrons. Core-electrons are highly localized and close to the nuclei. They rarely participate in interatomic bonding and resist changes in chemical environment, whereby maintaining their atomic characteristics. Thus, they are often treated as fixed background charges, with their effects on valence-electrons encoded in the pseudopotentials which are transferable between chemically different systems. Pseudopotential-DFT seeks variational solutions solely based on valence-electron density. Not surprisingly, discussion of its accuracy is focused primarily on valence-electrons. However, such a practice does not preclude the influence of core-electron density, as the construction of the pseudopotentials inherently involves simulating core-electron potentials and the polarization effects of a physical all-electron atom. Error in core-electron density can translate into pseudopotential errors, subsequently affecting valence-electron density and band structure/properties. In other words, core density errors directly, although implicitly, influence the accuracy of solid-state DFT through pseudopotentials.

To understand how errors in core-electron density propagate into solids and how to mitigate them, knowing the correct density of solid is of paramount importance. For atomic species, high-level wavefunction methods, such as the all-electron coupled cluster singles and doubles approach, provide a nearly exact electron density distribution \cite{bochevarov_densities_2008}. This for solids is, however, a formidable challenge, as a universal ``gold standard'' is not yet established, although high-level periodic methods such as coupled-cluster and quantum Monte Carlo are increasingly providing reference-quality data for selected systems\cite{GruneisJCTC, Booth}. Thus, the correctness of an electronic density must be inferred through an implicit comparison with experiments, based on a specific benchmark such as bandgap or lattice constant. However, fulfilling a single benchmark may be achieved not from a correct electron density but from an error cancellation \cite{hammes-schiffer_conundrum_2017,medvedev_density_2017,graziano_quantum_2017}. If more than one physical quantity is used as the benchmark across a class of systems, then one can trust that the method, to a good extent, is based on the correct electron density. The central objective of DFT is to find the minimum energy $\textit{E}(\textit{N}_{0},\textit{V}_{0})$ for a system with $\textit{N}_0$ electrons occupying volume $\textit{V}_0$ at T = 0 K. The discontinuity in the derivative $\partial E/\partial N|_{N_0}$ yields the bandgap \cite{perdew_physical_1983,sham_density-functional_1983}, while the equilibrium condition $\partial E/\partial V|_{V_0}$ defines the equilibrium volume and thus the lattice constants. So, simultaneously benchmark bandgaps and lattice constants are perhaps the most natural way to measure success for a method.

In this work, we investigate how errors in Cu core-electron density affect bandgap and lattice constants, as a representative study of the effects on semi-core 3d electrons. Despite being featured by a simple 3d$^{10}$4s$^1$ valence-electron configuration with an experimental d-s gap, which should be suited for conventional DFT calculations, Cu-containing closed-shell semiconductors are among the most problematic systems for DFT; even hybrid functionals and GW turn out to be unsatisfactory \cite{zhang_electronic_2025,rasander_density_2013,shi_structural_2013,lukashev_electronic_2007,gao_quasiparticle_2018}. Notably, all these studies employ pseudopotentials generated by (semi-)local functionals. Here, we show that unphysical self-interactions in (semi-)local functionals lead to an overly delocalized density distribution and a significant overestimation of quantum screening effect of core-electrons (defined below) on nuclei-valence attraction. The errors are subsequently encoded in pseudopotentials and transferred to solid calculations, causing too high 3d energy positions. By realizing the importance of using different functionals for core/valence regions, consistent with pseudopotential partitioning, we achieve high-accuracy in both bandgaps and lattice constants for a whole class of over 50 Cu-containing semiconductors.

\section{METHOD AND Computational details}
Pseudopotential-DFT calculations were performed using the Quantum ESPRESSO package \cite{giannozzi_advanced_2017}. To highlight the role of Cu's core electrons, three norm-conserving pseudopotentials, generated by OPIUM using different functionals, local density approximation (LDA) and Perdew-Burke-Ernzerhof (PBE) generalized gradient approximation, and modified Hartree-Fock (mHF), were employed \cite{noauthor_see_nodate-1}. The mHF pseudopotential was generated from the hybrid functional that combines HF exchange with PBE correlation, i.e., \begin{equation}\label{(1)}
	E_{xc}^{mHF} = E_{x}^{HF} + E_{c}^{PBE},
\end{equation}
following Refs. \onlinecite{tan_understanding_2019,yang_hybrid_2018}. Unless otherwise specified, the pseudopotential for Cu has a 3d$^{10}$4s$^1$ valence-electron configuration. For other atoms, only the LDA/PBE pseudopotentials from the SG15 pseudopotential library were considered\cite{Schlipf2015SG15}. For valence-electrons, we used either LDA \cite{ceperley_ground_1980} or PBE \cite{perdew_generalized_1996} functional.

Hereafter, we use the short notation: pseudopotential@functional, e.g., mHF@LDA indicates that mHF pseudopotential is used for Cu, while LDA pseudopotential is used for other atoms, combined with the LDA exchange-correlation functional to describe the valence electrons in calculations. Note that the logic for pairing up two different functionals for core and valence electrons, mHF@PBE or mHF@LDA lies in that, in the core, electrons are highly localized with substantial self-interaction error \cite{wing_band_2021}, while in the lattice, electrons are delocalized and slowly varying. Hence, exchange-correlation functionals for them should be qualitatively different. This approach can also be viewed as an extension of the subsystem functional approach, originated by Mattsson and Kohn \cite{kohn_edge_1998,mattsson_energy_2001}.

\begin{figure*}
	\includegraphics[width=\textwidth]{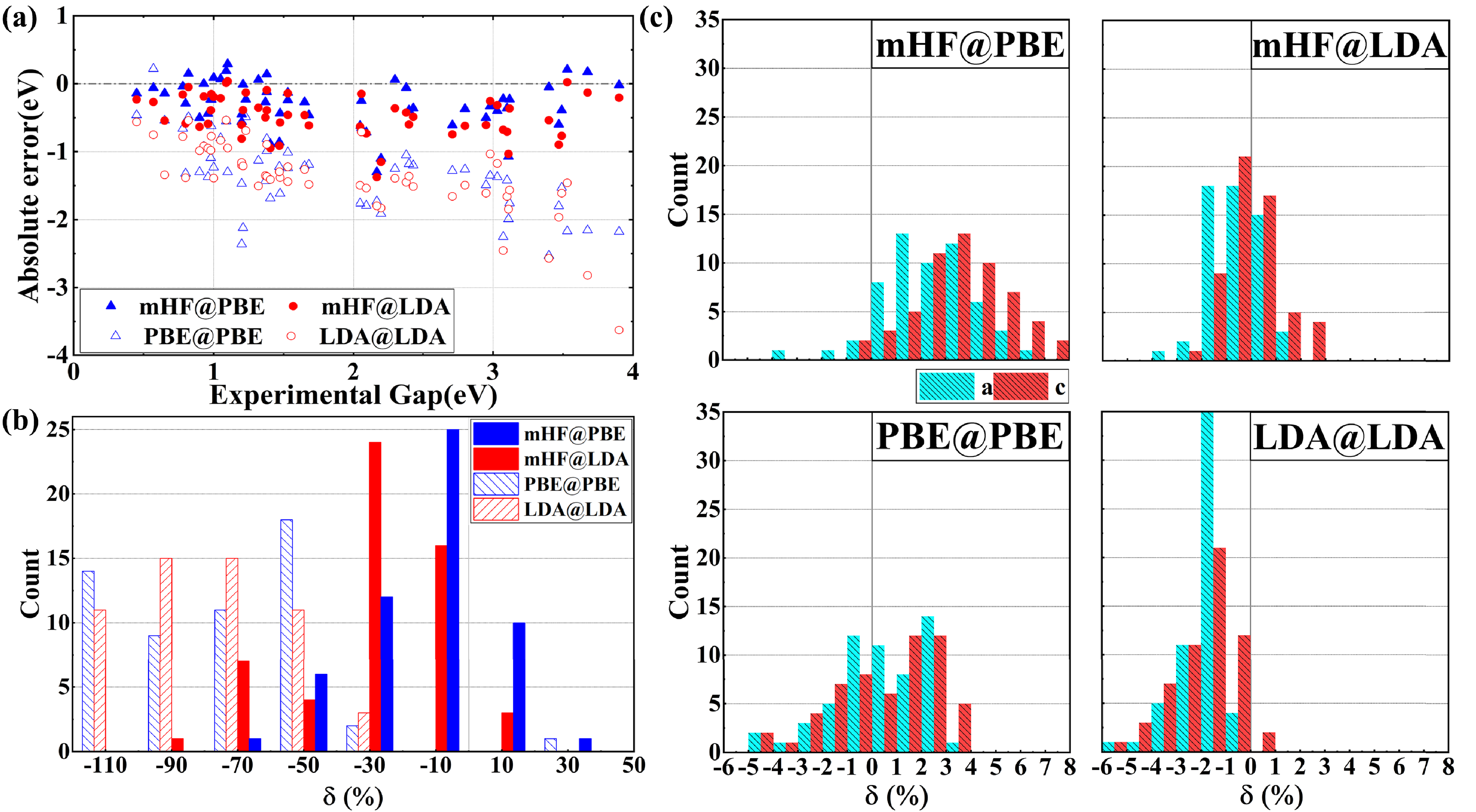}
	\caption{\label{fig:fig1} (a) Absolute errors of bandgap by four different methods for 54 Cu-containing closed-shell semiconductors relative to experimental values. When the system is erroneously metallic, the gap is defined between Cu-4s and -3d states at $\Gamma$ point. As observed in Supplementary Materials\cite{noauthor_see_nodate}, these negative bandgaps are mostly quite small, resulting in a marginal statistical impact on the mean relative error. When calculating the root-mean-square error, metallic systems are counted with a zero bandgap. For delafossite transparent conductive oxides, CuAlO$_2$, CuGaO$_2$ and CuInO$_2$, optical bandgaps are compared. (b) Probability distribution of errors for bandgap: $\delta (\%)= 100\times(E_g^c-E_g^e )/(E_g^e)$  where $E_g^c$ and $E_g^e$ are calculated and experimental bandgap, respectively. The histograms are at (0, $\pm$10\%, $\pm$30\%, etc.); each covers a range of $\pm$10\%. Also, for clarity they are not placed exactly on the spots. $\delta < -$100\% implies erroneous metal. (c) Probability distribution of errors for lattice constants $a$ and $c$ of 57 Cu-containing semiconductors, with histograms showing counts within an error interval. Note that the discrepancy between using 54 or 57 compounds for calculating bandgaps and lattice constants stems from the availability of experimental data, rather than any selective exclusion of computational results. Specifically, for Cu$_2$GeS$_3$ and Cu$_2$CdGeS$_4$, only experimental bandgap values are available, with no reported lattice constants. In contrast, for SrCuTeF, Cu$_2$CdGeTe$_4$, Cu$_2$CdSnTe$_4$, Cu$_2$ZnGeTe$_4$, and Cu$_2$ZnSnTe$_4$, only experimental lattice constants are available, with no bandgap data found. Details are given in Supplementary Materials \cite{noauthor_see_nodate}.}
\end{figure*}	

\section{RESULTS AND DISCUSSION}

This section consists of four subsections, organized as follows. In Subsection \textbf{A}, we compare performances of different pseudopotential-DFT calculations for the bandgap and lattice constant of monovalent-Cu semiconductors. In Subsection \textbf{B}, we show how different pseudopotentials affect the 3d and 4s charge densities and energy levels of Cu atoms. In Subsection \textbf{C}, we elaborate on how the charge-density and energy-level corrections for the Cu atoms are encoded into the mHF pseudopotential to propagate into solids. In Subsection \textbf{D}, we discuss the positioning philosophy for our use of pseudopotentials in this work and its practical applicability to other transition metal compounds.

\subsection{Monovalent-Cu semiconductors}

Figure 1 summarizes the performances of mHF@LDA, mHF@PBE, LDA@LDA, and PBE@PBE on bandgaps and lattice constants (See Supplementary Materials for details\cite{noauthor_see_nodate}). Qualitatively, only mHF yields positive bandgaps with no erroneous metal, regardless valence exchange-correlation functional. Quantitatively, the mean relative error (MRE) for bandgap is 20\% for mHF@PBE and 29\% for mHF@LDA. Their corresponding root-mean-square errors (RMSE) are 0.45 and 0.56 eV, respectively. In contrast, both PBE@PBE and LDA@LDA show considerable systematic underestimation of the bandgap and erroneous metal: 14 cases for the former and 11 cases for the latter. MREs (RMSEs) are also substantially larger, 80\% (1.35 eV) for PBE@PBE and 82\% (1.45 eV) for LDA@LDA, greatly limiting the practicality of these methods.

In their seminal work, Perdew et al. \cite{perdew_physical_1983} and Sham et al. \cite{sham_density-functional_1983} revealed that LDA/GGA underestimates bandgap due to their lack of derivative discontinuity. However, we find that mHF@LDA and mHF@PBE can substantially reduce the underestimation and in some cases, the results even approach the experimental bandgaps. Our understanding is that the change from the LDA/PBE pseudopotential to the mHF amounts to a drastic alteration of the effective external potential on the valence electrons\cite{shen_way_2024}. This alters the wavefunctions of valence states, thereby causing a significant change in the Kohn-Sham eigenvalues of the frontier states and, consequently, altering the bandgap. In the past, DFT development has long been focused on employing more advanced functionals for valence-electrons while maintaining LDA/GGA pseudopotentials. Our results, however, suggest that pseudopotentials should also be included in this pursuit in a holistic manner, as it not only accelerates DFT calculations but also provides a different perspective for correcting the bandgaps.

Although both produce acceptable bandgaps, mHF@PBE and mHF@LDA yield markedly different lattice constants, with the former severely overestimating them [see Fig. 1(c)]. For example, MRE for $a$ and $c$ is 2.9 and 3.7\%, respectively, in mHF@PBE with the maximum overestimation in $c$ $\sim$8\%. As shown in Supplementary Materials\cite{noauthor_see_nodate}, out of the 57 Cu-semiconductors studied, only 9 exhibit a lattice constant $a \neq b$. Our analysis indicates that the MREs for $b$ by different DFT calculations are nearly identical to those for $a$; therefore, for simplicity, we will not list statistics for $b$ separately hereafter. Note that 1-2\% error is the recognized performance of DFT \cite{mazdziarz_uncertainty_2024,yuk_putting_2024,haas_calculation_2009}. On the other hand, with MRE = 0.9\% for both $a$ and $c$, mHF@LDA here does an exceptionally good job. Its data points, with a $\sim$3\% maximum deviation, are tightly and nearly uniformly distributed around experimental values, exhibiting neither systematic underestimation nor overestimation. In contrast, although PBE@PBE, with MRE = 1.5\% for $a$ and 1.7\% for $c$, and LDA@LDA, with MRE = 2.0\% for $a$ and 1.9\% for $c$, are both good methods for lattice constant, PBE@PBE systematically overestimates the lattice constant, while LDA@LDA systematically underestimates it (see Fig. S3). Both have a maximum deviation of $\sim$5\%. Overall, mHF@LDA is statistically the best choice for lattice constants.
\begin{figure*}[htp]
	\includegraphics[width=\textwidth]{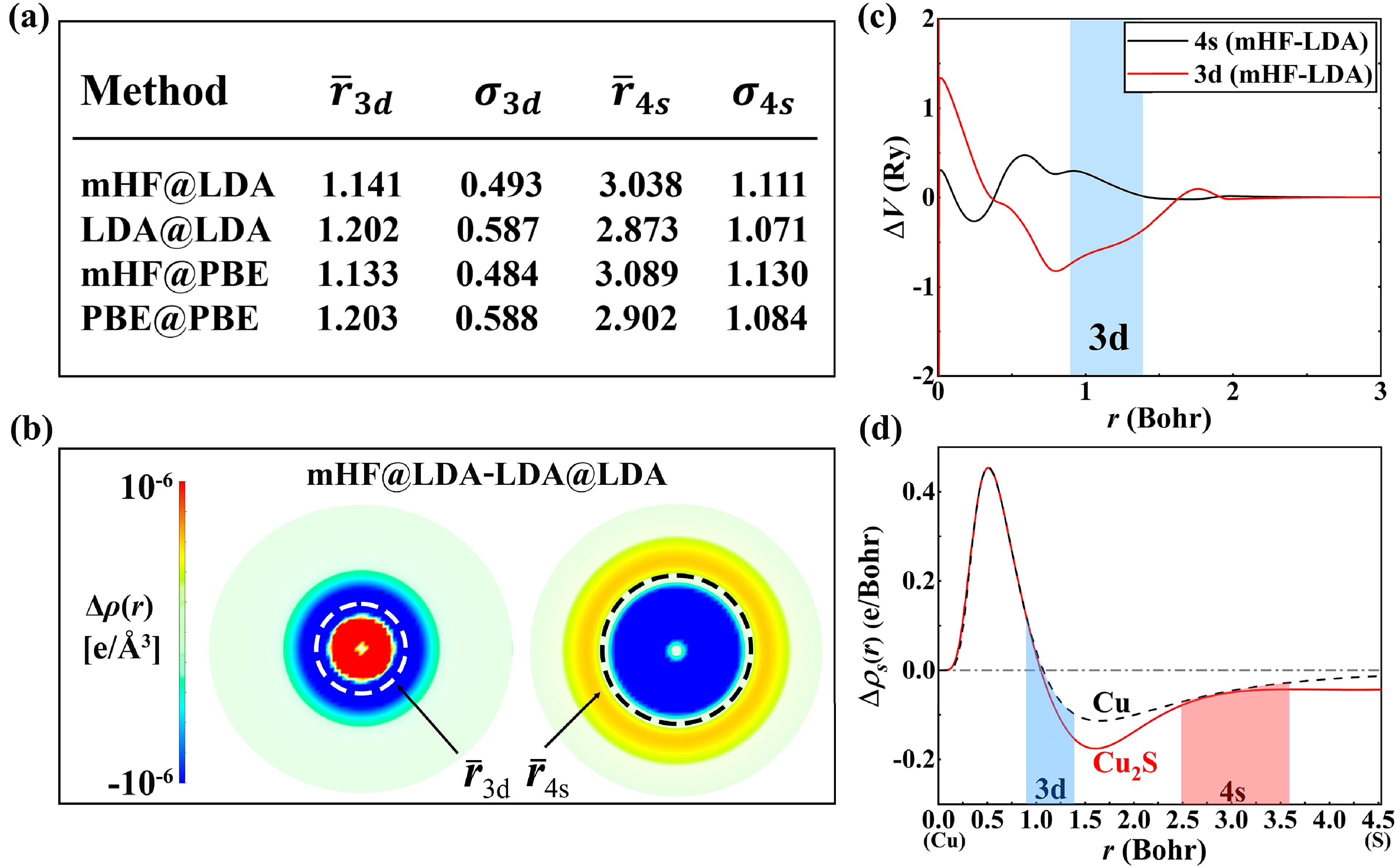}
	\caption{\label{fig:fig2} (a) Tabulation of 3d/4s orbital radius $\overline{r}$ and standard deviation $\sigma$ (both in Bohr) for Cu atom. (b) Density difference, $\Delta\rho(r) = \rho_{mHF}(r) - \rho_{LDA}(r)$ for Cu atom where $\rho_{mHF}(r)$ and $\rho_{LDA}(r)$ are mHF@LDA and LDA@LDA results, respectively, for 3d (left) and 4s (right) electrons. White and black dashed rings indicate positions of $\bar{r}_{3d}$ and $\bar{r}_{4s}$ by mHF@LDA in (a), respectively. (c) Differences in non-local pseudopotential for Cu 3d and 4s valence-electrons. Blue shading indicates the region centered at $\bar{r}_{3d}$ with a width of $\pm$0.5$\sigma_{3d}$ by mHF@LDA. (d) Differences in electron surface density between mHF@LDA and LDA@LDA, $\Delta\rho_s(r)$, for atomic Cu and bulk Cu$_2$S. Blue (red) shading indicates the region centered at $\bar{r}_{3d}$ ($\bar{r}_{4s}$) with a width of $\pm$0.5$\sigma_{3d}$ ($\pm$0.5$\sigma_{4s}$) by mHF@LDA for atomic Cu.}
\end{figure*}

\subsection{Cu atom}

Key to the performance difference revealed by above solid-state calculations lies in the various pseudopotentials, which encode the variations in the distribution of core-electron density. Since the pseudopotential is derived from the Cu atom, we proceed to investigate its effect on atomic charge densities and energy levels. To examine the importance of the core-electron density and its error propagation into solid calculations, we define
\begin{equation}\label{(2)}
	\begin{aligned}
		\overline{r}_i = \frac{\int r\rho_i(r)\mathrm{d}^3r}{\int \rho_i(r)\mathrm{d}^3r};\; \overline{r_i^2} = \frac{\int r^2\rho_i(r)\mathrm{d}^3r}{\int \rho_i(r)\mathrm{d}^3r};\; \mathrm{and} \,\quad \sigma_i = \sqrt{\overline{r_i^2}-\bar{r_i}^2},
	\end{aligned}
\end{equation}
where \textit{r} is the radial distance from the nucleus and $\rho_i(r)$ is the density distribution of Cu-3d and -4s electrons, respectively. The density-weighted $\bar{r}_i$ is a measure of the orbital radius, while the standard deviation $\sigma_i$ is a measure of the spread of $\bar{r}_i$. Figure 2(a) shows the results, from which one sees that mHF pseudopotentials produce consistently both smaller $\bar{r}_i$ and $\sigma_i$ for 3d electrons, indicating that the incorporation of exact exchange to core-electrons causes them to shrink toward the nucleus and become more localized. To balance the overall electron density, the 4s electrons move away from the nucleus and become more delocalized. Figure 2(b) depicts differential densities between mHF@LDA and LDA@LDA for Cu 3d and 4s electrons, respectively, where we see visually that opposite orbital displacements with respect to the nucleus have taken place. The decrease/increase in orbital radii cause a decrease/increase in 3d-4s energy splitting ($\Delta_{sd}$), which is increased from 0.74 (0.63) eV for LDA@LDA (PBE@PBE) to 4.35 (4.72) eV for mHF@LDA (mHF@PBE). Compared to the experimental value of 5.04 eV \cite{mann_configuration_2000}, mHF exhibits an order-of-magnitude improvement than both LDA and PBE, suggesting that it may indeed have a more accurate electron density.

In pseudopotential-DFT, pseudopotentials serve as the external potential for valence electrons \cite{shen_way_2024}. A too small $\Delta_{sd}$ implies that the pseudo core, LDA@LDA or PBE@PBE, severely differs from the true core. According to Hohenberg and Kohn \cite{hohenberg_inhomogeneous_1964}, ground-state electron density is uniquely determined by this external potential. Errors in external potential inevitably lead to errors in valence-electron density; the latter causes a series of erroneous predictions for solid state. In contrast, the pseudo core developed from mHF closely captures reality, thereby yielding better results as shown in Fig. 1.

Figure 2(c) compares the pseudopotential difference, $\Delta{V}$, between mHF and LDA for a Cu core, i.e., its nucleus plus 1s2s2p3s3p electrons, acting on 3d and 4s valence electrons. A negative (positive) $\Delta{V}$ indicates that mHF produces a stronger attraction (repulsion) than LDA. Near $\bar{r}_{3d}$ (shaded area), $\Delta{V}$ for 3d is negative, so mHF binds the 3d electrons more tightly to result in lower energy. Conversely, $\Delta{V}$ for the more delocalized 4s is overwhelmingly positive, so mHF pushes the 4s electrons farther away from the nucleus to result in higher energy.

\subsection{Atom-solid propagation}

\textit{Propagation of core-electron errors to solid via pseudopotentials.} To obtain quantitative insights across different chemical environments, it is desirable to remove the $(\theta,\psi)$-dependence of $\rho(r,\theta,\psi)$. To this end, we define the radial electron number as $n(r)=\int_{0}^{r}\int_{0}^{\pi}\int_{0}^{2\pi}\rho(r,\theta,\psi)r^2\sin\theta \mathrm{d}r\mathrm{d}\theta \mathrm{d}\psi$, and its radial derivative $\mathrm{d}n(r)/\mathrm{d}r=4\pi r^2\overline{\rho(r)}\equiv\rho_s(r)$, which represents the electron surface density over a spherical shell of radius $r$. Figure 2(d) compares the differences between mHF@LDA and LDA@LDA, termed $\Delta{\rho_s(r)}$, for Cu atom (black dashed curve) and for Cu$_2$S solid (red solid curve), respectively. A positive $\Delta{\rho_s(r)}$ implies a larger density of mHF@LDA than that of LDA@LDA, and vice versa. For $r\le1$ Bohr, $\Delta{\rho_s(r)}$ for atom and solid are nearly identical, implying that mHF pseudopotential has faithfully encoded the density of Cu core. For $r>1$ Bohr, on the other hand, $\Delta{\rho_s(r)}$ for atom and solid start to differ from each other. The difference is easily noticeable, as mHF affects the Cu$_2$S bonding, which in turn affects the electronic density distribution of the S anion - a point which will be revisited later when discussing lattice constants. From Fig. 2(d), one may attribute the much better Cu-3d level positions and opening of Cu$_2$S bandgap, opposed to be metallic (as predicted by LDA@LDA), to the transfer of core effects into solid by mHF pseudopotential. For a long time, consensus in the DFT community regarding this issue has been the misplacement of semi-core 3d energy \cite{karanovich_electronic_2021,tan_understanding_2019,wu_accuracy_2021,tan_effect_2018,oba_point_2011,lim_angle-resolved_2012}, not knowing that the root cause of the underestimation by LDA@LDA lies in the misplacement of core electron density.

\textit{How does a pseudopotential affect 3d semi-core electrons?} Today's prevailing understanding of pseudopotentials is that a wavefunction orthogonality requirement guarantees an effective repulsive force between electrons, which largely cancels the nuclear Coulomb attraction, leaving valence electrons subject to only a weak ``pseudopotential''. Here, we define this quantum effect as an effective ``screening'' by core electrons, as opposed to the classical electrostatic screening. In a way, pseudopotential represents the residual nuclear attraction after the quantum screening. When constructed using different functionals, the pseudopotential encodes different screening effects of core electrons to nuclear charge. The actual pseudopotential is generated from a functional approximation designed for a specific density distribution: e.g., LDA for a uniform electron gas, which accurately describes slow-varying near-free electrons \cite{kohn_edge_1998,mattsson_energy_2001}. In contrast, core electrons are tightly bound to the nucleus with both high density and large density variation, its behavior should be fundamentally different from free electrons. Moreover, due to unphysical self-interactions, LDA over-delocalizes the core-electron density, which erroneously enhances their screening effect and pushes, in turn, the semi-core 3d electrons further away from the nucleus to result in unrealistically high energy. When erroneous atomic density is used to construct LDA/PBE pseudopotentials, the too high 3d energy is transferred to that of solid. In contrast, mHF pseudopotentials employ exact exchange, which largely eliminates self-interactions and its associated wavefunction delocalization and over-screening. Consequently, mHF provides significantly improved 3d energy and bandgap.

\textit{Effect on lattice constants.} Lattice constants are primarily influenced by bonding: Stronger bonds result in shorter bond lengths and smaller lattice constants, and vice versa. LDA@LDA overestimates binding energy and thus systematically underestimates lattice constants. The mHF for core, relative to LDA, affects bonding in two ways: on the one hand, it leads to a decrease of Cu electron density in the bonding region, as evidenced by the negative $\Delta{\rho_s(r)}$ of Cu$_2$S in Fig. 2(d) in the intermediate region between Cu and S, e.g., $r\in(1.5,3.0)$ Bohr. Furthermore, $\Delta{\rho_s(r)}$ for Cu$_2$S lies below that for Cu atom implies that, in addition to Cu valence electron withdraw, some S valence electrons also move back towards S nucleus. Our calculations of S-centered $\Delta{\rho_s(r)}$ (not shown) confirm this. The physics behind all these changes lies in the redistribution of Cu 3d electrons due to the use of mHF pseudopotential. In other words, the propagation of Cu core electron density errors not only affects Cu d-s energy splitting (as discussed above) but also profoundly affects valence electron distribution even for the counter ions, i.e., S in Cu$_2$S. A reduction in the number of bonding electrons leads to a smaller binding energy (our calculations indeed show that mHF@LDA yields a smaller binding energy than LDA@LDA) and subsequently a longer bond length, which explains why in Fig. 1 mHF generally produces larger lattice constants compared to either LDA or PBE. On the other hand, as mHF increases spatial separation between Cu 3d and 4s electrons, leading to an outward movement of the latter [see Fig. 2(b)], followed by a significant reduction of the nuclear attraction, LDA, accurate for free electron gas, is therefore more appropriate for them \cite{ceperley_ground_1980}.

\subsection{Positioning and applicability}

Several points are worth noting. First, it is indeed the mHF pseudopotential that plays a decisive role, particularly in increasing the bandgap, rather than any shortcomings in the construction of LDA/PBE pseudopotentials. To this end, we have performed all-electron LDA calculations on representative semiconductors Cu$_2$S and CuCl using the Elk code\cite{ElkCode}, which yields bandgaps of 0 and 0.49 eV, respectively. As a comparison, results from LDA@LDA are 0 and 0.83 eV, and those from mHF@LDA are 0.82 and 2.86 eV, while the experimental values are 1.21 and 3.40 eV. Unambiguously, mHF constitutes the essential component for the improvement.

Second, as mentioned in the introduction, the bandgap and lattice constant are two of the most critical parameters in DFT outputs, and simultaneously achieving high-accuracy for both is perhaps one of the most straightforward ways to verify the success of a method. It is precisely because of their importance that such a wealth of experimental data is available as benchmarks. Although the focus here is on the bandgap and lattice constant, this does not imply that the effects of mHF are limited to these two aspects. For example, just to name a few, a shift in the Cu-3d energy level inevitably alters the d-band center, affecting predictions of catalytic performance; changes in the interatomic charge distribution influence chemical bonding, affecting formation energies; and the widening of the bandgap directly alters defect levels and exciton binding energies. In Fig. 1, the optical bandgap is used as a comparison for delafossite transparent conductive oxides, CuAlO$_2$, CuGaO$_2$ and CuInO$_2$, which serves as a ready-made example.

Third, it is imperative to clarify the positioning for our use of pseudopotentials in this work. Since the seminal work of norm-conserving pseudopotentials by Hamann, Schl\"{u}ter, and Chiang\cite{Hamann1979NormConserving}, pseudopotentials have generally been regarded as a mathematical tool for accelerating electronic structure calculations to reproduce all-electron results. However, we note that pseudopotentials originated in the 1930s and were used in electronic structure calculations\cite{hellmann_new_1935,cohen_fifty_1983} even before the advent of the Hohenberg-Kohn theorem that laid the foundation for DFT. As an effective potential, pseudopotentials are diverse in both their construction and application. Some pseudopotentials are constructed empirically, with the goal of matching experimental results; others are constructed theoretically, with the goal of matching first-principles results\cite{Melius1974EffectivePotentials}. The present work takes a different positioning of pseudopotentials: while they are constructed theoretically, the goal is no longer to reproduce first-principles all-electron results, but rather to serve as a means to correct errors in standard DFT, thereby obtaining results that better agree with experimental data. A direct consequence of this shift in positioning is that the choice of pseudo core significantly influences the results.
	
As explained earlier, the LDA/GGA error for the Cu atom and its monovalent semiconductors manifests as excessive screening of the nuclear attraction by the inner-shell 1s2s2p3s3p electrons. When all these electrons are included in the pseudo core to generate self-interaction-free mHF pseudopotential, the overestimation of the Cu-3d energy level is fully corrected. If just a subset of inner-shell electrons (e.g., 1s2s2p) is included in the pseudo core, the excessive screening would only be partially corrected because the self-interactions of the 3s3p electrons still persist. Consequently, the inward motion of the 3d electrons is hindered to some extent, preventing a complete correction of the energy-level overestimation. This would translate to a $\Delta_{sd}$ lying between that of conventional DFT and that of the mHF with a 1s2s2p3s3p pseudo core.

\begin{table}[htbp]
		\centering
		\caption{A comparison of the bandgaps and lattice constants for four representative Cu-semiconductors, calculated using the mHF-17@LDA and mHF-11@LDA methods, with their experimental values. Cu$_2$S and CuI have cubic structures, with lattice constants $a = b = c$, while CuAlO$_2$ and CuGaS$_2$ have the delafossite and chalcopyrite structures, respectively, with lattice constants $a = b$.
		}
		\label{tab:mhf_17e_11e}
		\small
		\setlength{\tabcolsep}{5pt}
		\renewcommand{\arraystretch}{1.18}
		
		\setlength{\arrayrulewidth}{0.5pt}
		\begin{tabular*}{0.95\textwidth}{@{\extracolsep{\fill}}llccccccc@{}}
			\hline
			\multirow{2}{*}{System}
			& \multirow{2}{*}{mHF}
			& \multicolumn{3}{c}{Bandgap (eV)}
			& \multicolumn{2}{c}{Calculated lattice (\AA)}
			& \multicolumn{2}{c}{Experimental lattice (\AA)} \\
			\cline{3-5} \cline{6-7} \cline{8-9}
			&
			& $E_g^{\mathrm{calc}}$
			& $E_g^{\mathrm{exp}}$
			& $\Delta E_g$
			& $a$
			& $c$
			& $a$
			& $c$ \\
			\hline
			
			\multirow{2}{*}{Cu$_2$S}
			& 17e & 0 & \multirow{2}{*}{1.21} & $-1.21$ & 5.978 & -- & \multirow{2}{*}{5.629} & \multirow{2}{*}{--} \\
			& \textbf{11e} & \textbf{0.82} &  & \textbf{$-$0.39} & \textbf{5.534} & -- &  &  \\
			
			\addlinespace[2pt]

			\multirow{2}{*}{CuAlO$_2$}
			& 17e & 2.17 & \multirow{2}{*}{3.53} & $-1.36$ & 2.824 & 17.66 & \multirow{2}{*}{2.860} & \multirow{2}{*}{16.95} \\
			& \textbf{11e} & \textbf{3.55} &  & \textbf{+0.02} & \textbf{2.763} & \textbf{16.87} &  &  \\
			
			\addlinespace[2pt]
			
			\multirow{2}{*}{CuGaS$_2$}
			& 17e & 1.55 & \multirow{2}{*}{2.43} & $-0.88$ & 5.771 & 9.491 & \multirow{2}{*}{5.349} & \multirow{2}{*}{10.47} \\
			& \textbf{11e} & \textbf{1.94} &  & \textbf{$-$0.49} & \textbf{5.337} & \textbf{10.39} &  &  \\
			
			\addlinespace[2pt]
			
			\multirow{2}{*}{CuI}
			& 17e & 1.72 & \multirow{2}{*}{3.12} & $-1.40$ & 6.453 & -- & \multirow{2}{*}{6.052} & \multirow{2}{*}{--} \\
			& \textbf{11e} & \textbf{2.75} &  & \textbf{$-$0.37} & \textbf{6.074} & -- &  &  \\
			
			\hline
		\end{tabular*}
	\end{table}
	
Our numerical calculations confirm this. We construct a 17-electron mHF pseudopotential with the 1s$^2$2s$^2$2p$^6$3s$^2$ pseudo core, termed mHF-17. mHF-17@LDA yields a $\Delta_{sd}$ of 2.11 eV for the Cu atom, which is smaller than 4.35 eV by mHF-11@LDA but larger than 0.74/0.63 eV by LDA@LDA/PBE@PBE. Table I compares the bandgaps and lattice constants of four representative Cu-semiconductors obtained from mHF-17@LDA and mHF-11@LDA, along with the corresponding experimental values. Clearly, the magnitude of the corrections from mHF-17 is smaller than that from mHF-11.

Finally, we believe the finding in Cu is not a special case. Similar effects are expected in other transition metals, but the situation warrants consideration from two perspectives. On the one hand, it is expected that mHF causes d-electrons to contract toward the nucleus and undergo a consequent energy lowering, as this is the result of eliminating self-interactions. On the other hand, whether this leads to a systematic improvement in the bandgap and/or lattice constant depends on the underlying causes of DFT errors. If the d-band energy overestimation is the primary factor, an improvement similar to that observed in monovalent-Cu semiconductors is expected. If it is merely one of several important factors, an improvement will still occur, but its magnitude becomes limited. If it does not constitute a major factor, substantial improvement is unlikely.

\section{CONCLUSION}

In summary, we investigate the relationship between Cu 3d/4s electron density and various forms of exchange-correlation functionals for Cu atom and calculated the resulting errors in bandgap and lattice constants for over 50 Cu-containing simple closed-shell semiconductors. We show that bandgap is sensitive to the 3d electron density. Obtaining a better result requires an accurate description of the core-electron quantum screening. At the same time, lattice constants also depend on the 4s electron density and it requires an accurate description of both the core- and valence-electron densities to get them right. By partitioning real space using pseudopotentials into regions optimized for different functionals, mHF@LDA minimizes the overall core errors propagation, achieving high-accuracy in both bandgap and lattice constants simultaneously. It suggests that the form of exact exchange-correlation functional may be more complex, dictated by local electron density and its derivatives, $\partial\rho$, e.g., close to mHF in core but to LDA in valence region, rather than one single form fits all. In other words, in an all-electron hybrid approach, the mixing parameter $\alpha$ may be a function of $\rho$ and $\partial\rho$ with a step function- or Fermi-Dirac-like transition between the core and valence electrons.

\section*{supplementary material}
Tables and figures are compiled to summarize the bandgaps and lattice constants obtained from different computational approaches, alongside available experimental results for systematic comparison.

\begin{acknowledgments}
This work was supported by the Ministry of Science and Technology of China (Grant No. 2023YFA1406400) and the National Natural Science Foundation of China (Grant No. 12474064). SBZ has no financial interests in any of the funding mentioned above.
\end{acknowledgments}

\section*{Conflict of Interest}
The authors have no conflicts to disclose.

\section*{reference}

\bibliographystyle{apsrev4-1}
\bibliography{reference.bib}

\end{document}